\let\accentvec\vec
\documentclass{llncs}
\usepackage[pdftex]{graphicx}
\usepackage{booktabs}
\usepackage{array}
\usepackage{listings}
\usepackage[caption=false]{subfig}
\usepackage{epstopdf}

\let\vec\accentvec

\usepackage{amsmath}
\usepackage{amsfonts}
\usepackage{amssymb}
\usepackage{amsthm}
\usepackage{bbm}
\usepackage{amsbsy}
\usepackage{threeparttable}
\usepackage{hyperref}
\usepackage{color}
\definecolor{MyDarkBlue}{rgb}{0,0.08,0.50}
\definecolor{BrickRed}{rgb}{0.65,0.08,0}
\hypersetup{
   colorlinks=true,       
   linkcolor=MyDarkBlue,          
   citecolor=BrickRed,        
   filecolor=red,      
   urlcolor=cyan           
}

\begin{document}

\frontmatter         
\title{Improving the Performance of Trickle-Based Data Dissemination in Low-Power Networks}

\author{Milosh~Stolikj,
        Thomas~M.~M.~Meyfroyt,
        Pieter~J.~L.~Cuijpers,
        and~Johan~J.~Lukkien}
\institute{Dept. of Mathematics and Computer Science, Eindhoven University of Technology, \\
        P.O. Box 513, 5600 MB, Eindhoven, The Netherlands}

\maketitle

\begin{abstract}
Trickle is a polite gossip algorithm for managing communication traffic. It is of particular interest in low-power wireless networks for reducing the amount of control traffic, as in routing protocols (RPL), or reducing network congestion, as in multicast protocols (MPL). Trickle is used at the network or application level, and relies on up-to-date information on the activity of neighbors. This makes it vulnerable to interference from the media access control layer, which we explore in this paper. We present several scenarios how the MAC layer in low-power radios violates Trickle timing. As a case study, we analyze the impact of CSMA/CA with ContikiMAC on Trickle's performance. Additionally, we propose a solution called \textit{Cleansing} that resolves these issues.
\end{abstract}

\section{Introduction}
Low-power wireless networks, such as networks of ubiquitous sensors, are being built with the aim to be available for extended periods of time, while using as little energy as possible. This includes wireless sensor networks in forests for detecting fires, in pipelines for detecting leaks, on light poles along streets to control luminosity etc~\cite{Akyildiz2002393}. In such resource-constrained devices, wireless transmissions are the largest source of power consumption. Therefore, networking protocols for low-power wireless networks are designed to avoid unnecessary traffic, such as redundant control information, or to prevent broadcast storms.

Trickle~\cite{Levis2004} has been proposed as an efficient algorithm for controlling traffic flow. It is being used in routing protocols for reducing the amount of control traffic~\cite{ctp,rfc6550}, in multicast protocols for reducing redundant repetitions of data packets~\cite{draft-ietf-roll-trickle-mcast} and in software update algorithms for managing the propagation of updates~\cite{Levis2004}. Trickle uses two premises to achieve fast propagation and reduced traffic: (1) suppressed transmissions when consistent information has been recently propagated by neighboring nodes, and (2) dynamic transmission rates depending on the consistency of information in the network. The concept of consistency is left to the application layer, which allows the Trickle algorithm to be implemented in different protocols.

The Trickle algorithm relies on accurate timing information in order to work as designed. However, various factors can influence this timing and can cause inconsistencies within the protocol. External disturbances can come from the radio medium (packet loss), network (congestion) and locally (data link layer). In this work, we analyze how the media access control (MAC) layer of low-power radios influences broadcast-based data dissemination using Trickle. As a case study, we consider a MAC layer comprised unslotted Carrier Sense Multiple Access with Collision Avoidance (CSMA/CA) and radio duty cycling. We show that due to contended media and CSMA/CA introduced back-offs, nodes can be starved from Trickle updates. This results in large propagation delays and inefficient messaging, making Trickle unsuitable for deadline-critical applications.

We discuss and analyze two common scenarios where there is a large discrepancy between the measured and expected update delay of Trickle, caused by the MAC layer. To resolve this, we propose a modification to the MAC layer to support dropping of queued Trickle packets based on incoming Trickle packets, called \textit{Cleansing}. Using simulations and experiments we show that the Cleansing MAC modification drastically improves the update delay in bottleneck topologies, and helps reduce the number of transmissions in grid-like topologies.

The paper is structured as follows. First, we cover related work on Trickle in Section~\ref{sec:relatedwork}. Then, we introduce the Trickle algorithm and the low-power protocols at the MAC layer in Section~\ref{sec:trickle-controlled-algorithms}. Next, in Section~\ref{sec:interference}, we describe how the MAC layer violates Trickle timing, and analyze this unwanted behaviour in two topologies. Section~\ref{sec:newmac} introduces the Cleansing improvements to the MAC layer. Finally, we compare simulation and experimental results of Trickle with and without Cleansing support in Section~\ref{sec:results} and give concluding remarks in Section~\ref{sec:conclusion}.

\section{Related work}
\label{sec:relatedwork}
The Trickle algorithm has been initially designed as an efficient method to disseminate software updates in low-power networks~\cite{Levis2004}. However, since it only specifies \emph{when} messages should be sent, and not \emph{how}, it has been accommodated in many other protocols~\cite{Levis2008}, such as network reprogramming~\cite{Lin2008}, routing~\cite{ctp,rfc6550} and data dissemination~\cite{codedrip-ewsn2014}. Trickle was recently  standardized~\cite{rfc6206} and used as a basis for the Multicast Protocol for Low power and Lossy Networks (MPL)~\cite{draft-ietf-roll-trickle-mcast}.

Various aspects of the Trickle algorithm have been studied so far. For example, in~\cite{trickle-ewsn2014,Vallati2013}, Trickle has been observed as unfair in terms of load share - certain nodes transmit more often than others. Trickle in absence of a MAC layer has previously been analyzed, e.g.,~\cite{becker,kermajani2012,meyfroyt2014}. Similarly, CSMA/CA for low-power networks has been analyzed without considering the upper layers, e.g.,~\cite{dunkels2011b,nullduty}. Finally, the potential problematic interaction between Trickle-based data dissemination and radio duty cycling has been sketched in~\cite{narrowcast-ewsn2014}, along with potential energy efficiency improvements by reducing the scope of single-hop broadcasts. However, to the best of the authors' knowledge, a detailed analysis on the interaction between Trickle and the MAC layer, consisting of both CSMA/CA and radio duty cycling, their combined performance and potential problems in specific topologies, has not yet been conducted, which is what this paper aims to do. The analysis and the results presented in this paper explain the simulation results for MPL in~\cite{Clausen2013,Oikonomou2013}, and the poor performance for small Trickle interval lengths.
\section{Trickle-based protocols}
\label{sec:trickle-controlled-algorithms}
\begin{figure}[ht]
\centering
  \includegraphics[width=0.25\textwidth]{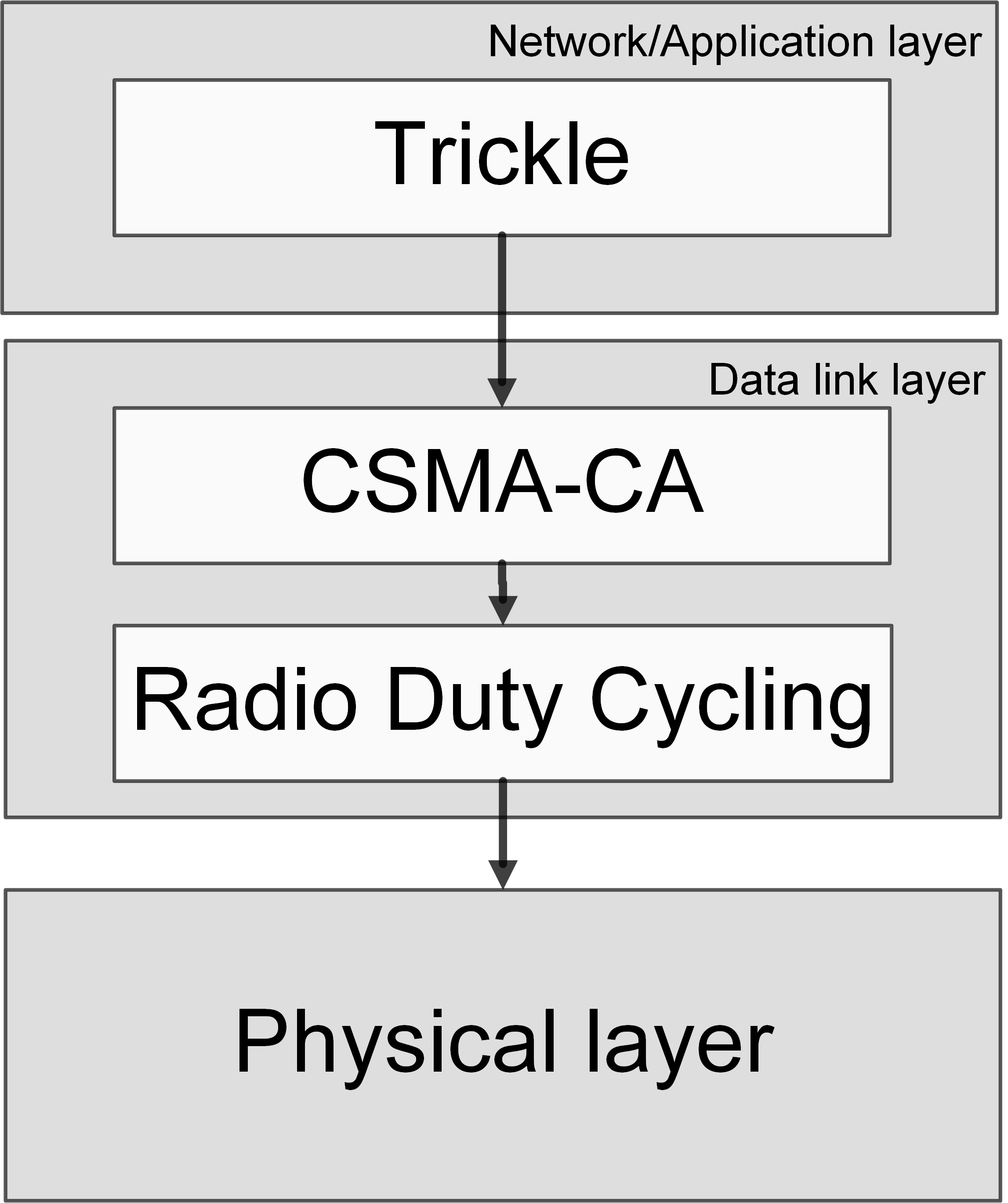}
\caption{Flow of Trickle packets in the Contiki operating system~\cite{dunkels04contiki}. }
\vspace{-2.0em}
\label{fig:layers}
\end{figure}

The Trickle algorithm is used mostly by communication protocols at the network or the application layer. Trickle essentially controls the generation of packets within these protocols. The lower layers are responsible for the actual transmission of the data packets sent by Trickle (Figure~\ref{fig:layers}).

The data link layer of low-power radios as IEEE 802.15.4~\cite{802.15.4e}, which is the focus in this work, is built of two components -  media access control (MAC) and a radio handling protocol. The MAC protocol handles the allocation of the shared medium among nodes and covers retransmissions in case of collisions or packet loss. The radio handling protocol determines the efficient use of the radio during the periods allocated by the MAC protocol.

We will now give a detailed description of the Trickle algorithm and the underlying MAC layer protocols.

\subsection{Trickle algorithm}
\label{sec:trickle}
Trickle has two main goals. Firstly, whenever new information becomes available in the network, it must be propagated quickly to all nodes. Secondly, when there is no update, communication overhead has to be kept to a minimum. The Trickle algorithm achieves this by moderating the number of packets that nodes generate with a ``polite gossip" policy.


We now provide a precise description of Trickle as it is given in~\cite{meyfroyt} (see also~\cite{Levis2004}). The algorithm has four global parameters, which are the same at every node in the network: a threshold value $k$, called the redundancy constant, minimum ($I_{\min}$) and maximum interval size ($I_{\max}$), and a listen-only parameter ($\eta$), which defines the size of a listen-only period. By default, $\eta=1/2$. Furthermore, each node in the network has its own timer and keeps track of three local variables: the size of the current interval ($I$), a counter ($c$) of the number of consistent data packets received during the current interval, and the transmission time ($t$) in the current interval.

The behavior of each node is described by the following set of rules. At the start of a new interval a node resets its timer and counter $c$ and sets $t$ to a value in $[\eta I,I]$ at random. When a node receives a new data packet that is consistent with the information it has, it increments $c$ by 1. When a node's timer reaches time $t$ and if $c < k$, it sends a data packet to its MAC layer queue. When a node's interval ends, it sets its interval size to $\min(2I,I_{\max})$ and starts a new interval. When a node receives a data packet that is inconsistent with its own information, then if $I > I_{\min}$ it sets $I$ to $I_{\min}$ and starts a new interval.

Trickle only determines when nodes should transmit; the nature of the transmission (broadcast/unicast), the structure of the message, and the exact definition of what is a consistent transmission is given by the upper layers, i.e. the protocols where Trickle is used. For instance, in dissemination protocols, as multicast, transmissions are always broadcasts; a node receives a consistent transmission when a known data packet is received from another node, and an inconsistent transmission is received when a new, unseen data packet is received.

\begin{figure}[ht]
\centering
  \includegraphics[width=0.7\textwidth]{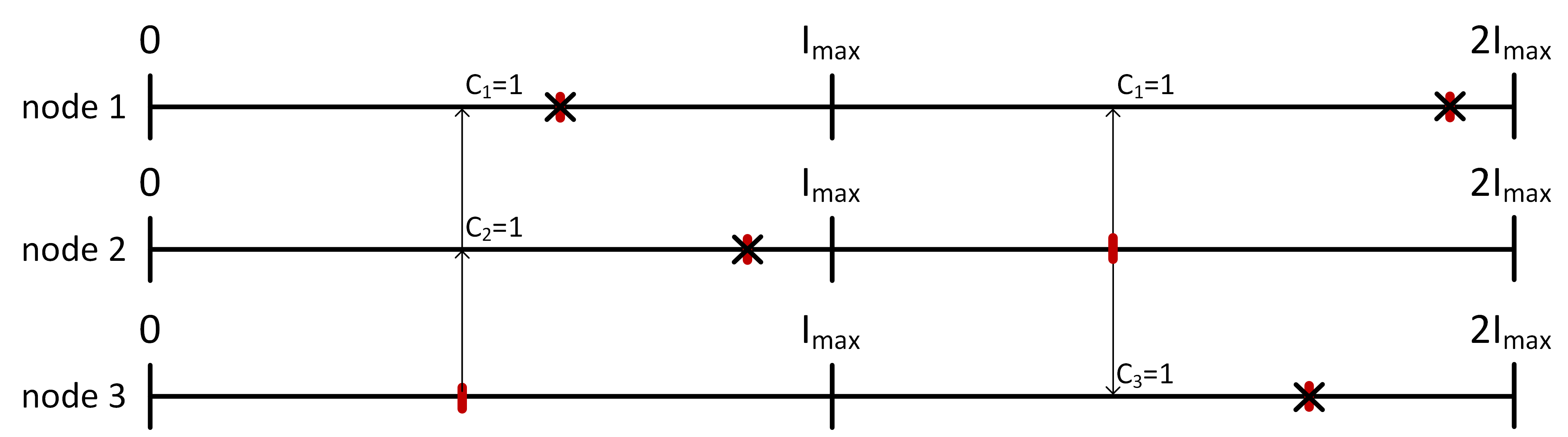}
\caption{Example of three synchronized nodes using the Trickle algorithm ($k=1$, $I=I_{\max}$). In the first interval, the transmissions by nodes 1 and 2 are suppressed by the transmission of node 3, while in the second interval, node 2 suppresses nodes 1 and 3.}
\vspace{-1.0em}
\label{fig:trickle}
\end{figure}


In Figure~\ref{fig:trickle} an example is depicted of a network consisting of three nodes using the Trickle algorithm with $k = 1$ and $I = I_{\max}$ for all nodes. Note that while in the example the intervals of the three nodes are synchronized, in general, the times at which nodes start their intervals need not be synchronized. In practice, networks will generally not be synchronized, since synchronization requires additional communication and consequently imposes energy overhead. Furthermore, as nodes get updated and start new intervals, they automatically lose synchronicity.

The four Trickle parameters can be used to tweak the algorithm behavior according to specific scenarios, giving option for trading between redundancy, speed of propagation and risk of collisions. For instance, $I_{\min}$ provides a trade-off between speed of propagation and number of packets: lower values of $I_{\min}$ will make nodes transmit sooner, though with an increased risk of collisions, and therefore, additional transmissions. To prevent such scenarios, the Trickle RFC recommends setting $I_{\min}$ to a multiple of the worst-case link layer latency, defined as the time until the first link-layer transmission of a frame, assuming an idle channel. Typical values of the Trickle parameters for various protocols are given in Table~\ref{table:parameters}. In the remainder of this paper, we will focus on broadcast-based data dissemination as the Trickle application protocol, similar to the MPL protocol, with the recommended value for the redundancy constant ($k=1$).

\begin{table}[ht]
\begin{center}
\begin{threeparttable}
\caption{Default values of Trickle parameters in different protocols.}
\vspace{-1.0em}
\begin{tabular}{ l r r r }
	\toprule
	Protocol               & $k$    & $I_{\min}$	& $I_{\max}$ \\ \toprule
    MPL (control traffic)~\cite{draft-ietf-roll-trickle-mcast}  & 1        & 10 times worst-case link-layer latency & 300 s \\
    MPL (data traffic)     & 1        & 10 times expected link-layer latency & $I_{\min}$ \\
    RPL (DIO)~\cite{rfc6550}       & 10       & 8 ms & 8.280 s \\
    CTP~\cite{ctp}           & $\infty (0)$       & 125 ms   & 500 s \\
\bottomrule	
\end{tabular}
\label{table:parameters}
\end{threeparttable}
\end{center}
\vspace{-2.0em}
\end{table}

\subsection{CSMA/CA protocol}
The actual transmission of packets generated by Trickle is left to the MAC layer. Protocols at this layer handle the allocation of the shared media among nodes and cover retransmissions in case of collisions or packet loss. The IEEE 802.15.4 MAC defines two flavors of the CSMA/CA protocol, depending on the operational mode in use: slotted CSMA/CA, used in beacon-enabled modes, where beacons are sent to synchronize nodes to a super-frame structure; and unslotted CSMA/CA, used in non beacon-enabled modes, where no beacons are sent out and there is no synchronization between nodes. In this paper, we focus on unslotted CSMA/CA, but the same concepts apply to slotted CSMA/CA.

In unslotted CSMA/CA, the basic time unit is the back-off period $\mathit{BP}$, which is related to the transmission time of a frame. Every node maintains two variables for each frame it wants to send: a back-off exponent $\mathit{BE}$, and a counter for the number of back-offs for the current transmission $\mathit{NB}$. These variables are controlled by three parameters: the minimum back-off exponent $BE_{\min}$, the maximum back-off exponent $\mathit{BE}_{\max}$ and the maximum number of back-offs $\mathit{NB}_{\max}$.

Initially, $\mathit{NB}=0$ and $\mathit{BE}=BE_{\min}$. Before each transmission, each node first waits for a random number of $BP$s ranging from 0 to $2^{\mathit{BE}}-1$. After the initial back-off, the node performs a clear-channel assessment (CCA) to determine whether the channel is free. If the channel is free, the node proceeds with the transmission. Otherwise, it increases $\mathit{NB}$ by one, and sets $BE$ to $\min(\mathit{BE}+1, \mathit{BE}_{\max})$. If $\mathit{NB} \leq \mathit{NB}_{\max}$, the entire procedure is repeated. After $\mathit{NB}_{\max}+1$ failed attempts, the frame is dropped from the MAC queue.

\subsection{Radio duty cycling}
The MAC layer of low-power radios often includes a second component next to the CSMA/CA protocol - the radio handling protocol. Radio transceivers are among the biggest sources of energy consumption in low-power wireless devices. Therefore, low-power wireless devices must trade-off between keeping the radio transceiver off, to save energy, and periodically wake up to be able to receive data from their neighbors. During the years, many radio duty cycling (RDC) protocols  have been proposed. They can be categorized into synchronous, where nodes are synchronized with their neighbouring nodes, and asynchronous, where no pre-synchronization is required. Asynchronous RDCs can be further categorized into sender initiated and receiver initiated protocols. Sender initiated RDC protocols give the transmission incentive to the senders: senders wake up receivers to receive a transmission. Receiver initiated protocols give the incentive to the receivers: receivers inform senders when they are prepared to receive a transmission. Finally, hybrid approaches have been developed, which combine features from any of the given categories.

\begin{figure}[ht]
\vspace{-1.0em}
\centering
  \includegraphics[width=0.6\textwidth]{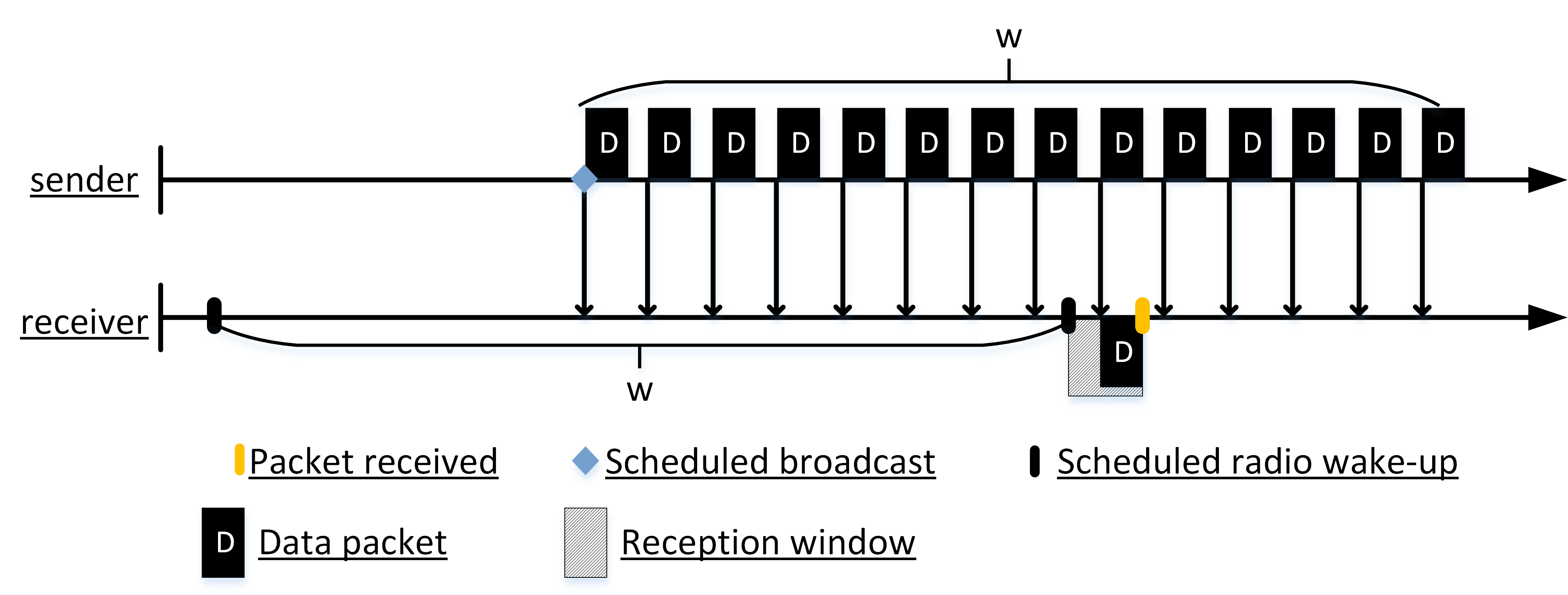}
\vspace{-0.5em}
\caption{In ContikiMAC, broadcast transmissions are sent with repeated frames for the full wake-up interval. This illustration is reproduced based on~\cite{dunkels2011b}.}
\label{fig:contikimac-broadcast}
\vspace{-1.0em}
\end{figure}

In this work, we consider ContikiMAC~\cite{dunkels2011b}, a sender initiated RDC. It is similar to the Coordinated Sampled Listening protocol (CSL), introduced in the IEEE 802.15.4e standard~\cite{802.15.4e}. A brief description of ContikiMAC follows.

By default, every node has its radio turned off. Periodically, at regular intervals of $w$ time units, each node turns its radio on to check for incoming traffic. If a transmission is detected, the radio is kept on until the frame is received. Transmissions are non-periodic, originating from the upper layer(s). When they arrive, a CCA is done to see if the medium is free. If it is free, the node starts transmitting immediately. Broadcast transmissions should be received by all nodes, irrespective of their wake up intervals. Therefore, a broadcast transmission will always be repeated for $w$ time units (Figure~\ref{fig:contikimac-broadcast}), so that each node will at least once turn on its radio during the transmission. Hence, assuming an idle channel, the worst-case latency as defined in the Trickle RFC, is $w$. However, this makes broadcasts expensive both in terms of delay and consumed energy.

The main configuration parameter for ContikiMAC is the radio wake-up frequency $1/w$, i.e. how often each node samples the radio. This parameter also dictates the maximum duration for each individual transmission $w$. Typically, the wake-up frequencies is set to 4Hz, 8Hz or 16Hz, giving wake up intervals of 250$ms$, 125$ms$ and 62.5$ms$, respectively. Reducing the wake-up frequency reduces the energy usage in the network, at the expense of a higher delay. 
\section{Interference scenario}
\label{sec:interference}
A common feature of both sender initiated and receiver initiated RDC protocols is that transmissions are not instantaneous, and there is a variable delay between the intent to start a transmission and the actual receipt. In sender initiated RDC protocols as ContikiMAC, the transmission starts almost immediately after it is received from the upper layers, but it is not completed until the receiver performs its periodic wake up to sample the channel. Similarly, in receiver initiated RDC protocols, the transmission is delayed until the sender receives a request from the receiver, which is again periodically scheduled. Finally, in case of collisions, in both cases, CSMA/CA will re-schedule transmissions after a certain back-off period. The delayed completion of a transmission creates a window where upper layer protocols may think that a transmission has been completed, while in fact, it is not. This causes unintended and inefficient messaging, as the transmission delay and retransmissions may move from one to another Trickle interval.

\begin{figure}[ht]
\centering
  \includegraphics[width=0.6\textwidth]{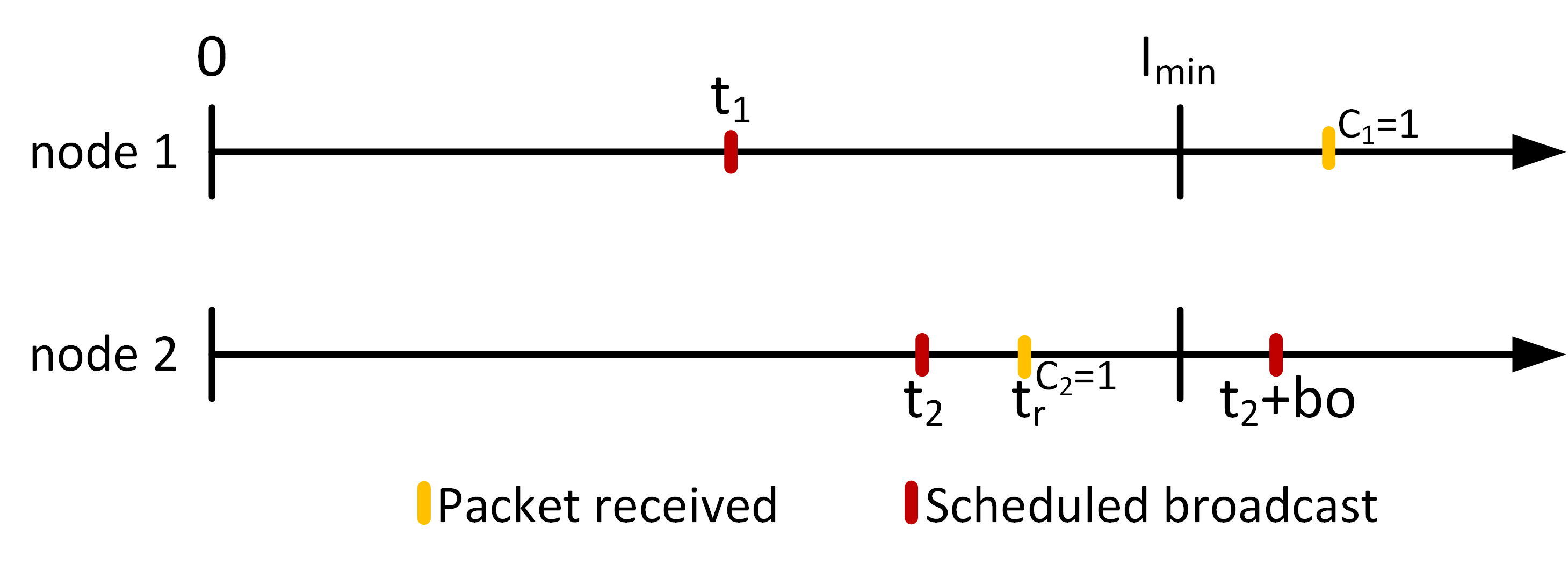}
\caption{MAC layer interference on Trickle timing. Nodes 1 and 2 get updated at the same time, and they select transmission times at $t_1$ and $t_2$, respectively. If the reception  for node 2 ($t_r$) is scheduled to be after $t_2$, node 2 will queue a Trickle packet at $t_2$, even though there is a packet in the air from node 1. Due to CSMA/CA, this packet will be transmitted after the back-off, at time $t_2+bo$.}
\label{fig:trickle-csma-rdc}
\vspace{-2.0em}
\end{figure}

For example, consider a network consisting of two nodes (Figure~\ref{fig:trickle-csma-rdc}). They use unslotted CSMA/CA in combination with radio duty cycling at the MAC layer. Packet transmission is regulated by the Trickle algorithm ($k=1$, $\eta=1/2$). Both nodes start a Trickle process at the same time, with consistent information for dissemination. They choose transmission times $t_1$ and $t_2$, respectively, such that $t_1 < t_2$. Both counters are initially set to zero ($c_1 = c_2 = 0$). At time $t_1$, since $c_1 < k$, node 1 sends a packet to its MAC layer. Then, it does a successful CCA and starts transmitting the packet. Node 2 has its next wake-up scheduled at time $t_r>t_2$. Consequently, at time $t_2$ node 2 has not yet received node 1's broadcast and will decide to transmit itself, sending a Trickle packet to its MAC layer. Since at this time the channel is busy, CSMA/CA will delay this transmission until $t_2 + bo$, where $bo$ is the back-off time. At time $t_r$, node 2 receives the transmission from node 1, setting $c_2 = 1$, making the queued packet in the MAC layer obsolete. However, since there is no link between the MAC queue and the application layer, the packet will be sent at $t_2 + bo$. This effect can be cascaded if multiple nodes exhibit the same behavior. Moreover, it is possible that node 2's broadcast is delayed into its next Trickle interval (Figure~\ref{fig:trickle-csma-rdc}), causing node 1 to suppress its next broadcast, further disrupting the Trickle process.

\subsection{Case study: CSMA/CA and ContikiMAC}
We will now use the Contiki operating system for a case study on the impact of MAC interference on Trickle timing. Contiki 2.7 utilizes the ContikiMAC RDC protocol with a radio wake-up interval length of $w$, together with a slightly modified version of the unslotted CSMA/CA protocol. Firstly, the default parameters $\mathit{BE}_{\min}=0$, $\mathit{BE}_{\max}=3$ and $\mathit{NB}_{\max}=3$, force CSMA/CA to skip the first back-off. Secondly, the back-off period is equal to the length of the wake-up interval of ContikiMAC ($\mathit{BP}=w$). As $w$ is the worst-case transmission time for ContikiMAC, this ensures that any retransmissions are attempted after the current transmission has finished. Thirdly, the CCA check is delegated to the RDC layer. Finally, the back-off exponent $\mathit{BE}$ is increased only when no acknowledgment is received for sent unicast frames. Since Trickle-based data dissemination uses only broadcast packets, for which no acknowledgment is needed, a back-off can only occur due to a failed CCA or a detected collision. In both cases, $\mathit{BE}$ remains one, causing the back-off for broadcasts to remain $\mathit{BP}=w$.

\subsubsection{Scenario 1: Single-hop network}
We now analyze the likelihood that the scenario discussed at the beginning of this section occurs under ContikiMAC. Denote by $\mathbb{P}^{\text{bo}}_{2}$ the probability that a CSMA back-off takes place in a network of two nodes.  For simplicity, we assume the nodes to be synchronized, which would be the case if they got updated simultaneously.  We assume that packets are received at radio wake-up and $I_{\min}=m \cdot w$, where $m\geq 2$ is a constant and $w$ is the radio wake-up interval. We require $m\geq2$, since otherwise a node will never be able to finish a transmission within the same Trickle interval as it was scheduled. Furthermore, assume that the Trickle process has $k=1$ and $\eta=1/2$. A CSMA back-off will take place if either node 1 or 2 pick their transmission time during a broadcast of the other node and before their radio wake-up and reception. Hence, we can write
\begin{equation}\label{p2b}
\mathbb{P}^{\text{bo}}_{2}:=2\mathbb{P}[t_1\leq t_2\leq t_r \leq t_1+w]
=2\int\limits_{I_{\min}/2}^{I_{\min}}\mathbb{P}\left[t_2\in[t_1,t_r]\text{ }\vert\text{ }t_1=t\right]\text{d}\mathbb{P}[t_1\leq t].
\end{equation}
Since both $t_1$ and $t_2$ are chosen uniformly in $[I_{\min}/2,I_{\min}]$ and a broadcast starting at time $t$ is received by the non-transmitting node uniformly at $t_r\in[t,t+w]$, some calculus gives
\begin{equation}\mathbb{P}^{\text{bo}}_{2}=\frac{2}{m}-\frac{4}{3m^2}.\end{equation}

Note that this probability only depends on $m$, the ratio between the length of an interval $I_{\min}$ and the length of a broadcast $w$. For the MPL standard $I_{\min}=10w$, this implies $\mathbb{P}^{\text{bo}}_{2}=0.1925$, which is relatively large.


Extending these calculations and noting that nodes choose their timers independently, the probability that $b$ CSMA back-offs occur and $b+1$ transmissions are scheduled during an interval in a single-hop network consisting of $n$ nodes is given by
\begin{equation}\label{pnbs} \mathbb{P}^{\text{bo}}_{n,b}:= n \binom{n-1}{b} \mathbb{P}\left[t_2\in[t_1,t_r]\right]^b\mathbb{P}\left[t_r\leq t_2\right]^{n-b-1}.
\end{equation}
Like \eqref{p2b}, this expression can be evaluated analytically and allows us to calculate the probability $\mathbb{P}^{\text{bo}}_n$ that at least one CSMA back-off ($b>0$) takes place during a single interval in a single-hop network consisting of $n$ nodes:
\begin{equation}\label{eq}\mathbb{P}^{\text{bo}}_{n}:=1-\mathbb{P}^{\text{bo}}_{n,0}=1-\frac{1}{m^n}\left((m-1)^n+\frac{1}{2n-1}\right).
\end{equation}
Moreover, calculating the expected number of redundant transmissions per interval due to poor interaction between Trickle and the CSMA protocol gives:
\begin{equation}\label{exps}\mathbb{E}[N^r_n]:=\sum_{i=0}^{n-1} i\mathbb{P}^{\text{bo}}_{n,i}= \frac{n}{m}-\frac{1}{n+1}\left(\frac{2}{m}\right)^n.\end{equation}
Hence, the expected number of obsolete broadcasts per interval due to timing issues grows linearly with the size of the single-hop broadcast range.  This is intuitive, since every node has the same probability of scheduling a back-off. If Trickle worked as designed, there would be only one packet per interval~\footnote{For a complete calculation of Equations (\ref{p2b}-\ref{exps}), see Appendix A}.

\subsubsection{Scenario 2: A bottleneck network}
\begin{figure}[ht]
\centering
  \includegraphics[width=0.4\textwidth]{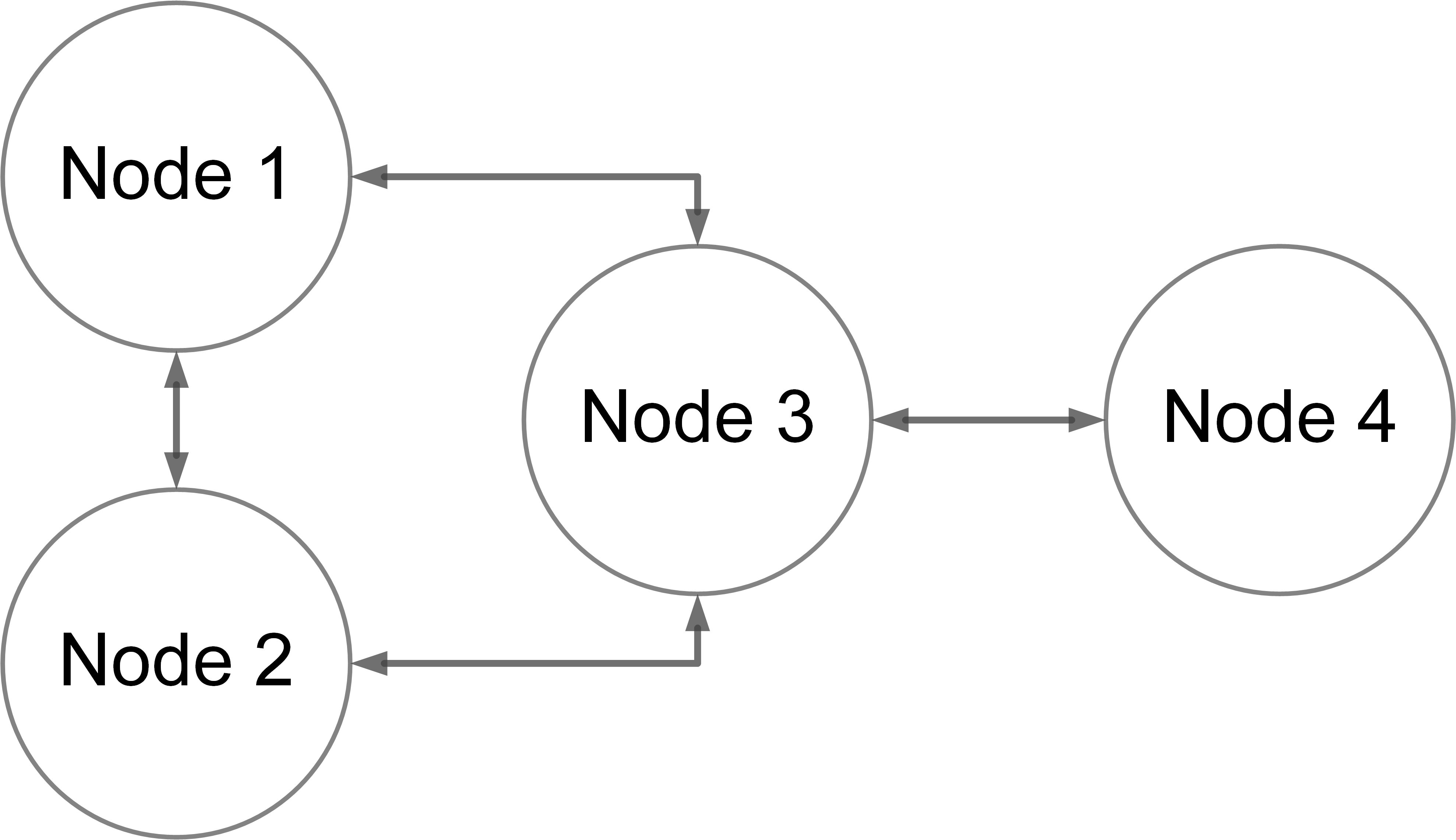}
\caption{A network consisting of 4 nodes, where node 3 is a bottleneck node.}
\label{fig:4nodes}
\vspace{-1.0em}
\end{figure}

Consider now a network of four nodes, with connectivity as in Figure~\ref{fig:4nodes}. This type of connectivity, where part of the network is reachable only through a single bridge node, is common, for example, in street lighting networks. Again all nodes use CSMA/CA in combination with ContikiMAC and run a Trickle dissemination process. The Trickle process has $k=1$, $\eta=1/2$ and $I_{\min}=m \cdot w$, where $m\geq2$ is a given constant. Initially, all nodes have consistent information and $I=I_{\max}$.

\begin{figure}[ht]
\centering
  \includegraphics[width=0.7\textwidth]{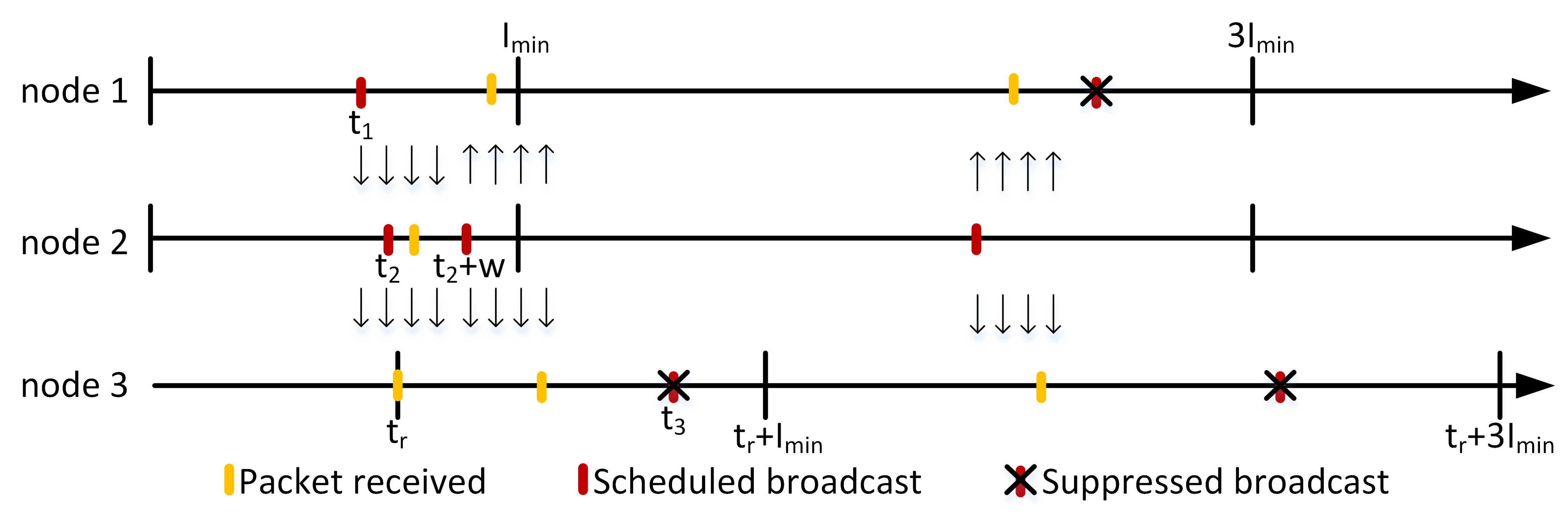}
\caption{Suppression of Trickle updates due to MAC layer interference. Nodes 1 and 2 get updated at the same time, and select transmission times at $t_1$ and $t_2$, respectively, with the periodic channel check for node 2 ($t_r$) scheduled to be after $t_2$. Node 2 queues a Trickle packet at $t_2$. Due to busy media, CSMA/CA re-schedules the packet for $t_2+w$. In the mean time, node 3 gets updated and starts a new Trickle interval. The re-transmission at $t_2+w$ causes node 3 to suppress its transmission in the first interval ($t_3$). As node 1 and 2 started the second interval earlier than node 3, there is a high probability that they will suppress any future transmissions from node 3.}
\label{fig:bottleneck_intervals}
\vspace{-1.0em}
\end{figure}

Suppose at time 0 nodes 1 and 2 receive an update simultaneously from a close-by source, set $I=I_{\min}$ and start a new interval (Figure~\ref{fig:bottleneck_intervals}). Node 1 is the first node to schedule a broadcast, which it starts to transmit at time $t_1$. As we have seen in the previous scenario, node 2 will schedule a broadcast before receiving node 1's broadcast with probability $\mathbb{P}^{\text{bo}}_{2}$. If this happens, the MAC protocol will cause node 2 to delay its transmission until time $t_2+w$. Before this time, however, node 3 will have been updated by node 1's transmission, and will start a new interval of length $I_{\min}$ and schedule a transmission at time $t_3$. Now node 2's transmission follows, suppressing node 3's transmission at time $t_3>t_2+w$ and consequently delaying the time that node 4 is updated. In its next interval, node 3 will broadcast only if it starts transmitting before it receives a broadcast by nodes 1 and 2. However, due to the synchronization caused by the Trickle protocol, this has a small probability, as can be seen in Figure~\ref{fig:bottleneck_intervals}. In the following intervals the same problem occurs. Only when node 4 eventually transmits its old information, which potentially could take a long time, it will reset node 3's Trickle process and an update will follow.

\begin{figure}[ht]
\vspace{-1.0em}
\centering
  \includegraphics[width=0.6\textwidth]{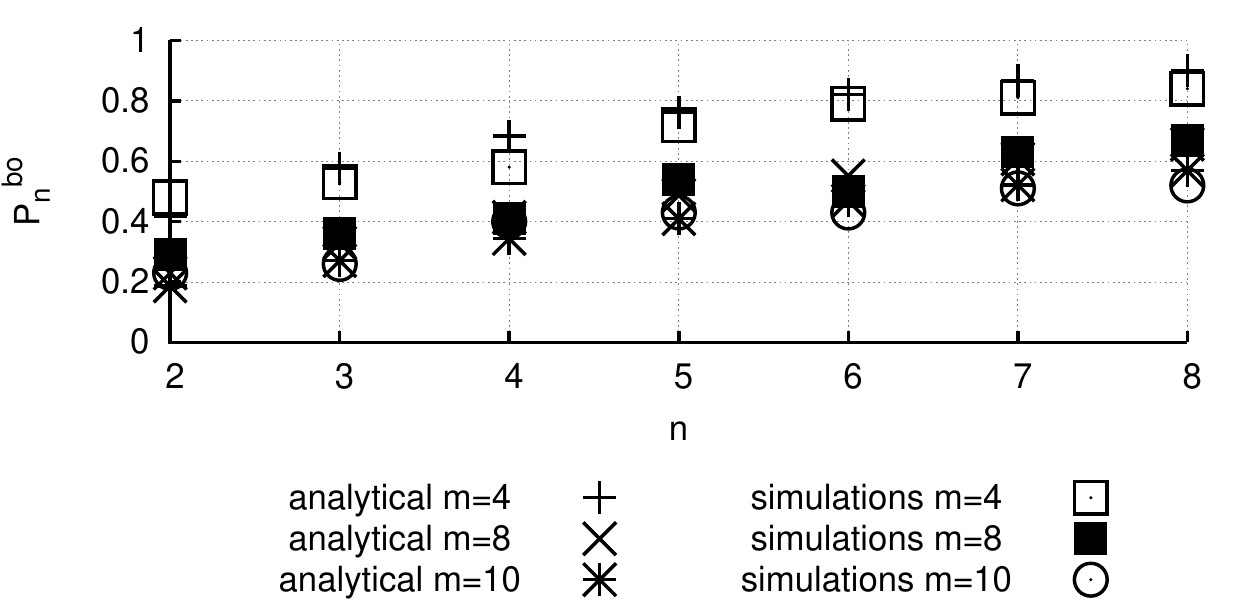}
\caption{Analytical and simulation results of the probability that node 4 is updated after the second Trickle interval, for different values of $m$ ($I_{\min}=m \cdot w$).}
\label{fig:pbo}
\vspace{-1.0em}
\end{figure}

In general, if node 3 is connected with $n$ synchronized nodes trying to update it, the previously described scenario occurs with probability $\mathbb{P}^{\text{bo}}_n$ (see \eqref{eq}). We have plotted this probability and compared it with simulations for different values of $m$ and $n$ in Figure \ref{fig:pbo}. From the plot it is clear that such an event is not rare. Given that such an event occurs, the probability that node 3 will ever broadcast in the following intervals before being suppressed by its neighbors is  small, even for $n=2$. Therefore, in such an event, with high probability node 4's update is delayed until it advertises its own old information, resetting the Trickle process of node 3. This gives an expected delay of approximately $\frac{1}{2}I_{\max}+\frac{3}{4}I_{\min}$, which is possibly very large since $I_{\max}$ is generally large. If node 4 has neighbors suppressing its own transmissions, then the expected delay will be even larger.

\section{Cleansing MAC}
\label{sec:newmac}
In order to reduce the interference of the data link layer on Trickle timing, we propose adding a \textit{Cleansing} mechanism to the MAC layer. If Trickle is treated as a network primitive, as suggested in~\cite{Levis2008}, known at both the network and data link layer, then some decision making can be done at the data link layer. Assuming that the MAC layer maintains separate queues per destination, whenever a new Trickle packet arrives from the network, the Cleansing MAC will purge any queued outgoing Trickle packets. This will lead to less redundant packets in the network, and will minimize the bottleneck problem from the previous section.

In most cases, purging outgoing Trickle packets improves Trickle performance in terms of messaging and delay, and does not lead to functional incorrectness. It remains consistent with the software design of low-power networks, as any purged packet can be seen as a message loss, and applications are already able to handle that situation. However, we can identify two scenarios where performance-wise, purging can be considered to be harmful.

The first scenario is when $k>1$, a purged Trickle message might not be obsolete. However, this should have minimal impact on the network, since only a small fraction of messages within each single-hop broadcast domain will be purged. Moreover, other nodes in reach will make up for the purged transmission.

The second scenario is when a Trickle message with an old value arrives, and the Cleansing MAC protocol purges an outgoing Trickle message with a new value, increasing the overall propagation delay. However, the effect of the purge is minimal, as due to the old message, the Trickle interval of the node with the new value will be set at $I_{\min}$, which would give a second opportunity for broadcast relatively soon. 
\section{Evaluation}
\label{sec:results}
To confirm the analytical results and to evaluate the performance of the Cleansing MAC modifications, we conducted several experiments in simulation and on a physical test bed. We used one application - dissemination of an update using Trickle, implemented in Contiki 2.7. Each experiment starts by injecting an update in the network. As the update is propagated, nodes increase their Trickle interval. The experiment ends when all nodes have reached their maximum Trickle interval $I_{\max}=10 \cdot I_{\min}$. We measured the delay, i.e. the time required to update all nodes, the total number of sent packets, the number of MAC layer retransmissions, and the mean waiting time in the MAC layer queue.

\subsection{Simulation results}
The simulations were carried out in the cross-level simulator Cooja~\cite{osterlind2006}. Cooja internally uses the MSPsim device emulator for cycle accurate Tmote Sky emulation~\cite{Polastre:2005:TEU:1147685.1147744}, as well as a symbol accurate emulation of the IEEE 802.15.4 CC2420 radio chip. We used the Unit Disk Graph Radio Medium propagation model, with no loss. All nodes use unslotted CSMA/CA with the default parameters ($\mathit{BE}_{\min}=0$, $\mathit{BE}_{\max}=3$, $\mathit{NB}_{\max}=3$), and the ContikiMAC RDC protocol, with a wake-up frequency of 8Hz ($w=125$ms). $I_{\min}$ varies from $250$ms to $1.75$s, at 250ms steps ($m=2, 4, ..., 14$), well beyond MPL's recommendation of $m=10$.

\subsection{Bottleneck topology}
The first scenario follows the bottleneck topology, as shown in Figure~\ref{fig:4nodes}. An update is inserted at the same time at nodes 1 and 2, and is propagated to the rest of the network using Trickle. Each configuration was simulated 1.000 times.
\begin{figure}[ht]
\vspace{-3.0em}
\centering
\subfloat[Trickle update interval and update delay]{
\centering
\includegraphics[width=.5\textwidth]{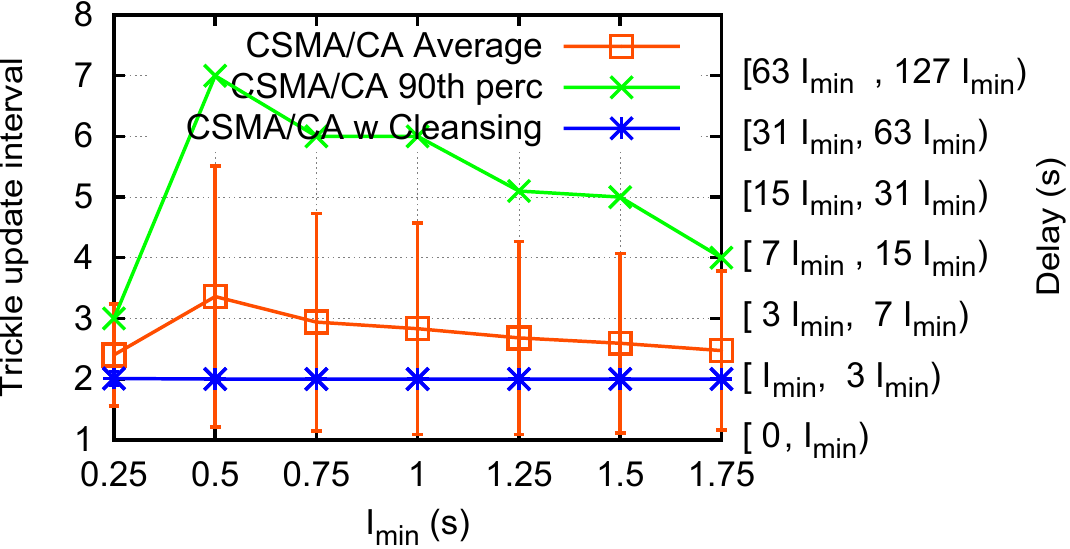}
\label{fig:4nodes-updateinterval}
}
\subfloat[Average update delay - worst 10\%]{
\centering
\includegraphics[width=.5\textwidth]{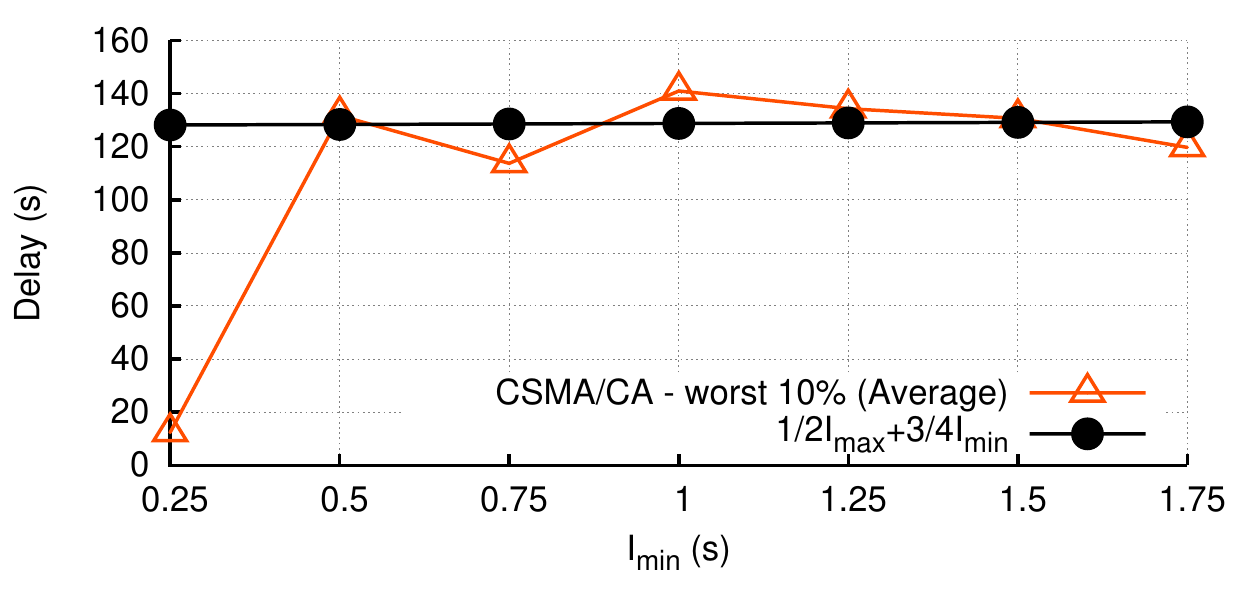}
\label{fig:4nodes-10thpercentile}
}
\caption{Update delay in the bottleneck scenario ($I_{\max}=256s$, $k=1$, $\eta=1/2$). a) shows the Trickle interval in which node 4 gets updated, with and without Cleansing MAC improvements. The left y axis shows the Trickle doubling interval, and the right y axis the actual time. b) shows the average delay of the largest 10\% of the measurements, and the analytical expected delay. The error bars correspond to the standard deviation.}
\vspace{-1.0em}
\end{figure}

As expected, without Cleansing, due to the large number of collisions, the update delay of node 4 is highly variable (Figure~\ref{fig:4nodes-updateinterval}). Both the mean and the standard deviation peak at $I_{\min}=0.5s$, and gradually decrease as $I_{\min}$ increases. Surprisingly, the update delay at $I_{\min}=0.25s$ is stable. This anomaly occurs because at $I_{\min}=0.25s=2 \cdot w$, the contention window of nodes 1 and 2 is equal to the broadcast duration ($w$). This practically guarantees collisions, and a retransmission from one of the nodes. However, node 3's listen-only period will be finished before the retransmission starts, and there is a chance that node 3 will schedule its own transmission before it receives the retransmission. Even if the transmission from node 3 is delayed, it will be sent within one or two broadcast periods. However, with $I_{\min}=0.5s$, nodes 1 and 2's contention window is still small, giving high probability for collisions. Then, retransmissions will always fall in node 3's listen-only period, forcing it to suppress its own transmission.

Figure~\ref{fig:4nodes-10thpercentile} depicts the average measured delay of the worst 10\% of the observations. This is a clear indication that harmful back-offs due to CSMA/CA are not uncommon, and that their effects can be detrimental to Trickle's performance. The update delay then becomes significantly high, in line with the analytical expected delay of $\frac{3}{4}I_{\min}+\frac{1}{2}I_{\max}$.

Finally, the interference is completely resolved with MAC Cleansing. In that case, updates are always completed in the second interval, as expected.

\subsection{Grid topology}
The second scenario consists of 100 nodes, arranged in a 10x10 grid, with 10 meters between two nodes in each axis. A new Trickle event is generated at the top left node. We simulate 100 executions of Trickle with different values for $I_{\min}$. Furthermore, we varied the connectivity range of each node. Each node has a circular coverage area with radius $2 + 10R$ meters, with $1\leq R \leq5$.

\begin{figure}[ht]
\vspace{-2.0em}
\centering
\subfloat[Update delay]{
\centering
\includegraphics[width=0.45\textwidth]{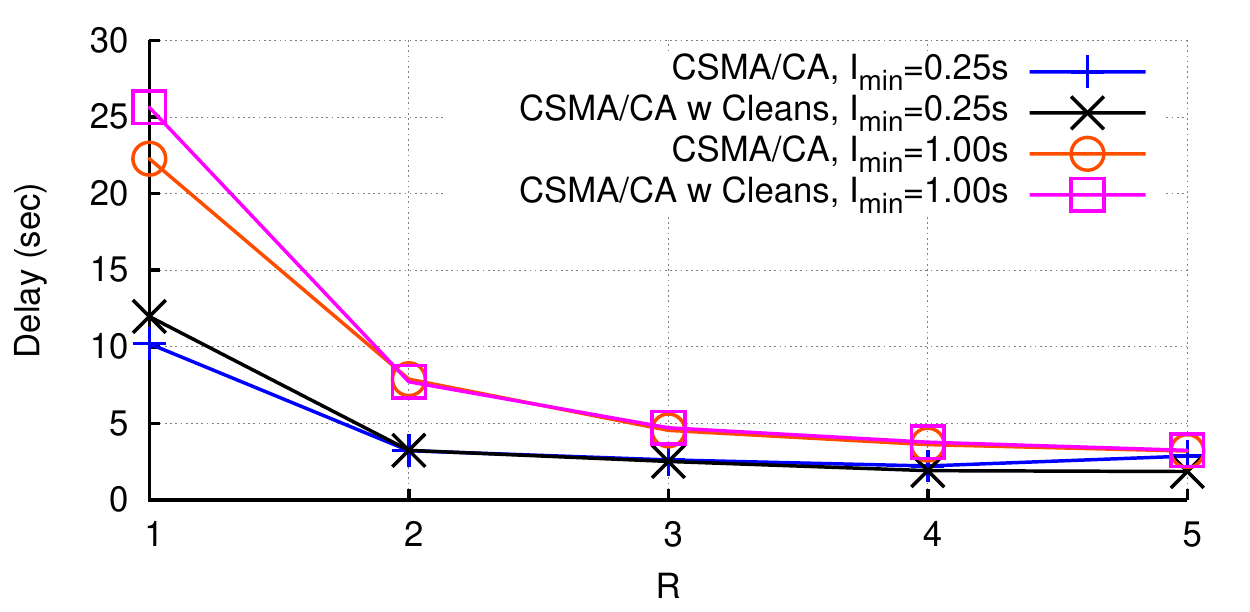}
\label{fig:delay-comparison}
}
\subfloat[Number of transmissions]{
\centering
\includegraphics[width=0.45\textwidth]{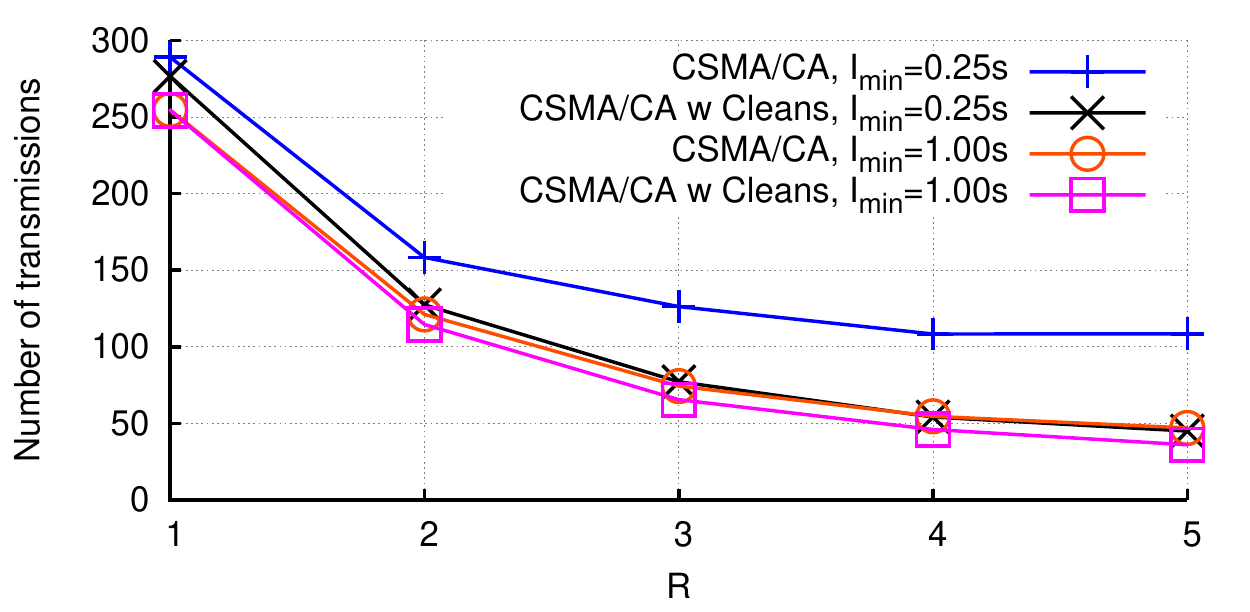}
\label{fig:tx-comparison}
}
\label{fig:comparison}
\vspace{-0.5em}
\caption{Average delay and average number of transmissions in the grid scenario. Using CSMA/CA with Cleansing with $I_{\min}=0.25s$ requires a similar number of transmissions as regular CSMA/CA with $I_{\min}=1.00s$, while the update delay is halved.}
\vspace{-1.5em}
\end{figure}
Figure~\ref{fig:delay-comparison} shows the update delay when using CSMA/CA with and without Cleansing. Since there are no bottlenecks in this scenario, these are comparable. However, the reduction in the number of sent packets is visible in Figure~\ref{fig:tx-comparison}. We can see that the number of transmissions with Cleansing is significantly lower than without Cleansing, while the average update delays are the same.

\begin{figure*}[h]
  \centering
    \vspace{-2.0em}
  \subfloat[Grid - TX]{
    \centering
  	\label{fig:gridtx}
  	\includegraphics[width=0.3\textwidth]{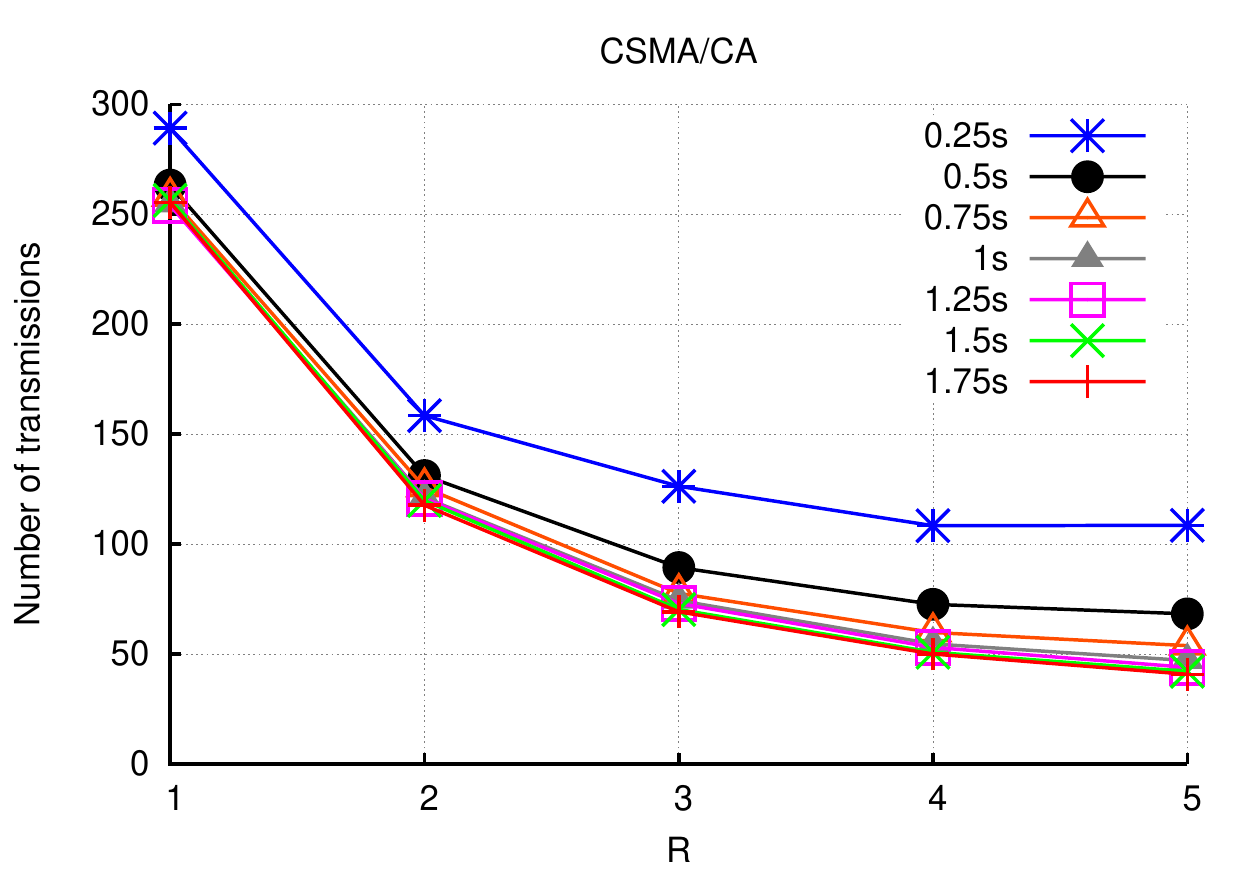}
  }
  \subfloat[Grid - RTX]{
    \centering
  	\label{fig:gridrtx}
  	\includegraphics[width=0.3\textwidth]{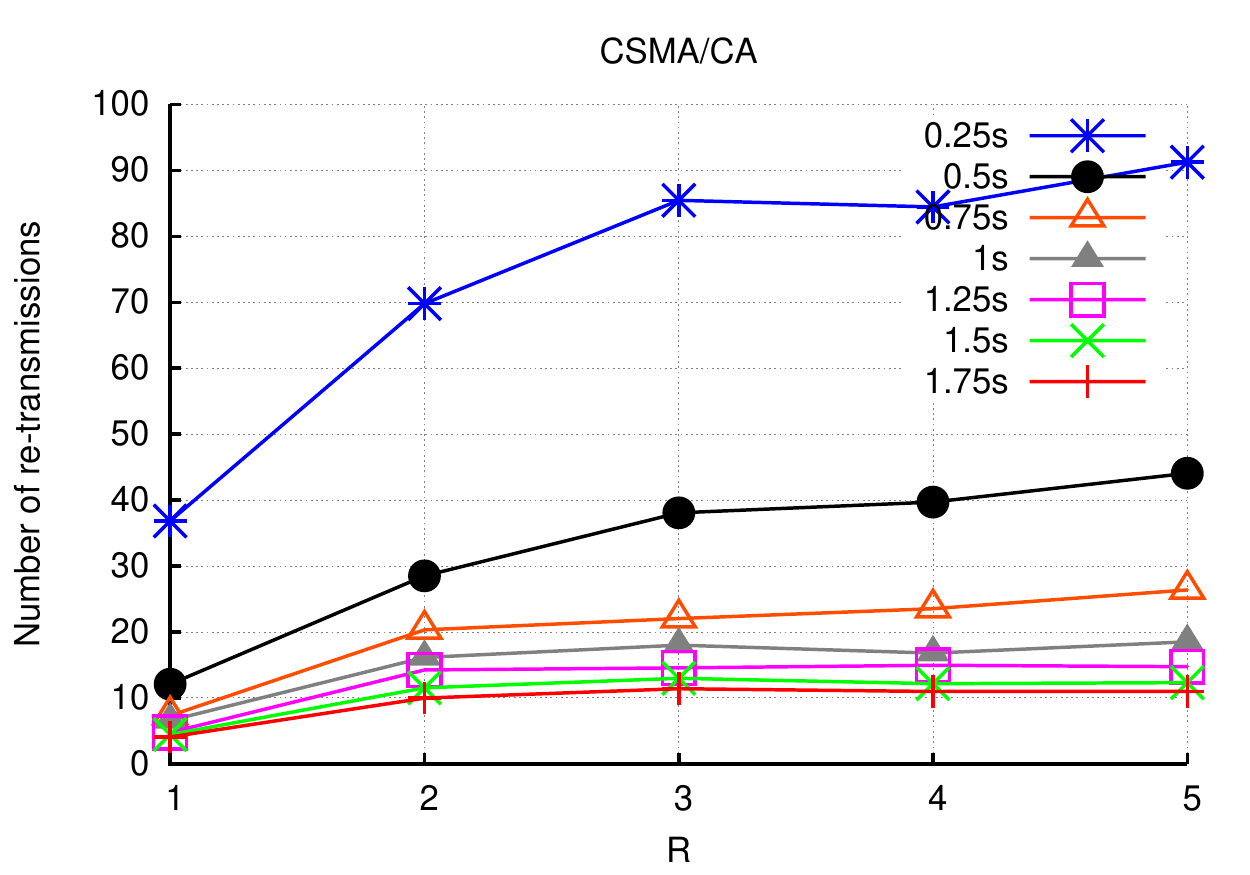}
  }
  \subfloat[Frame queue time]{
    \centering
  	\label{fig:gridqueue}
  	\includegraphics[width=0.3\textwidth]{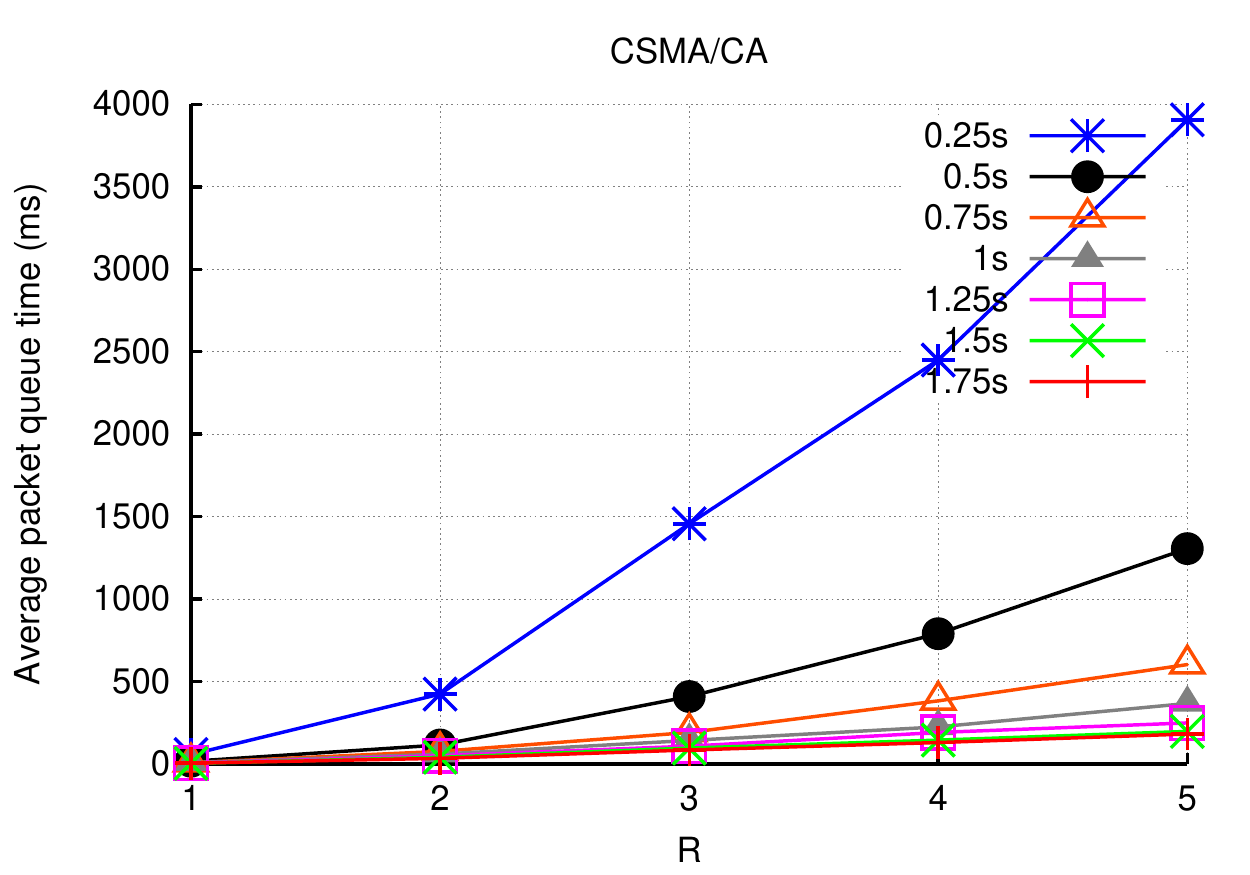}
  } \\\vspace{-1em}
  \subfloat[Grid - TX]{
    \centering
  	\label{fig:gridtx-cleans}
  	\includegraphics[width=0.3\textwidth]{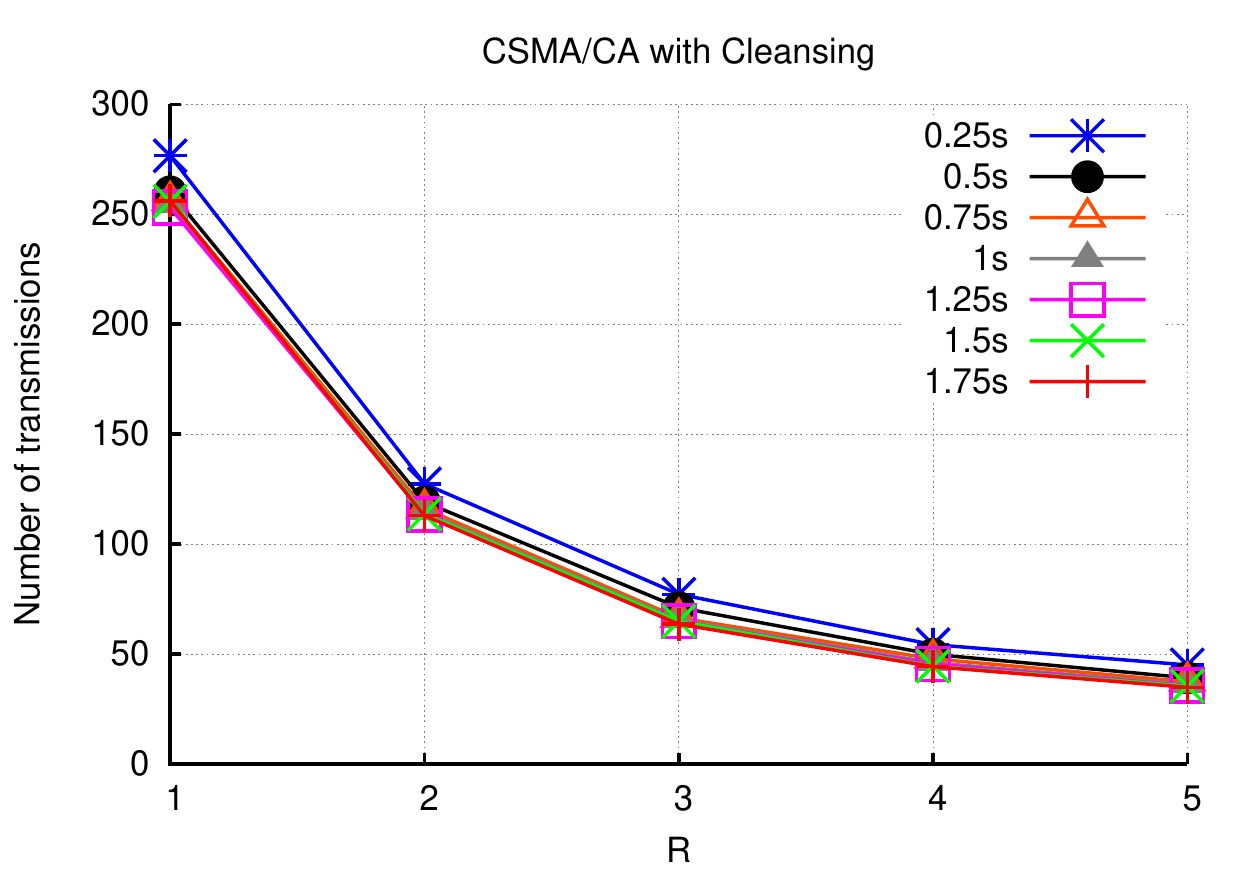}
  }
  \subfloat[Grid - Dropped frames]{
    \centering
  	\label{fig:griddrp-cleans}
  	\includegraphics[width=0.3\textwidth]{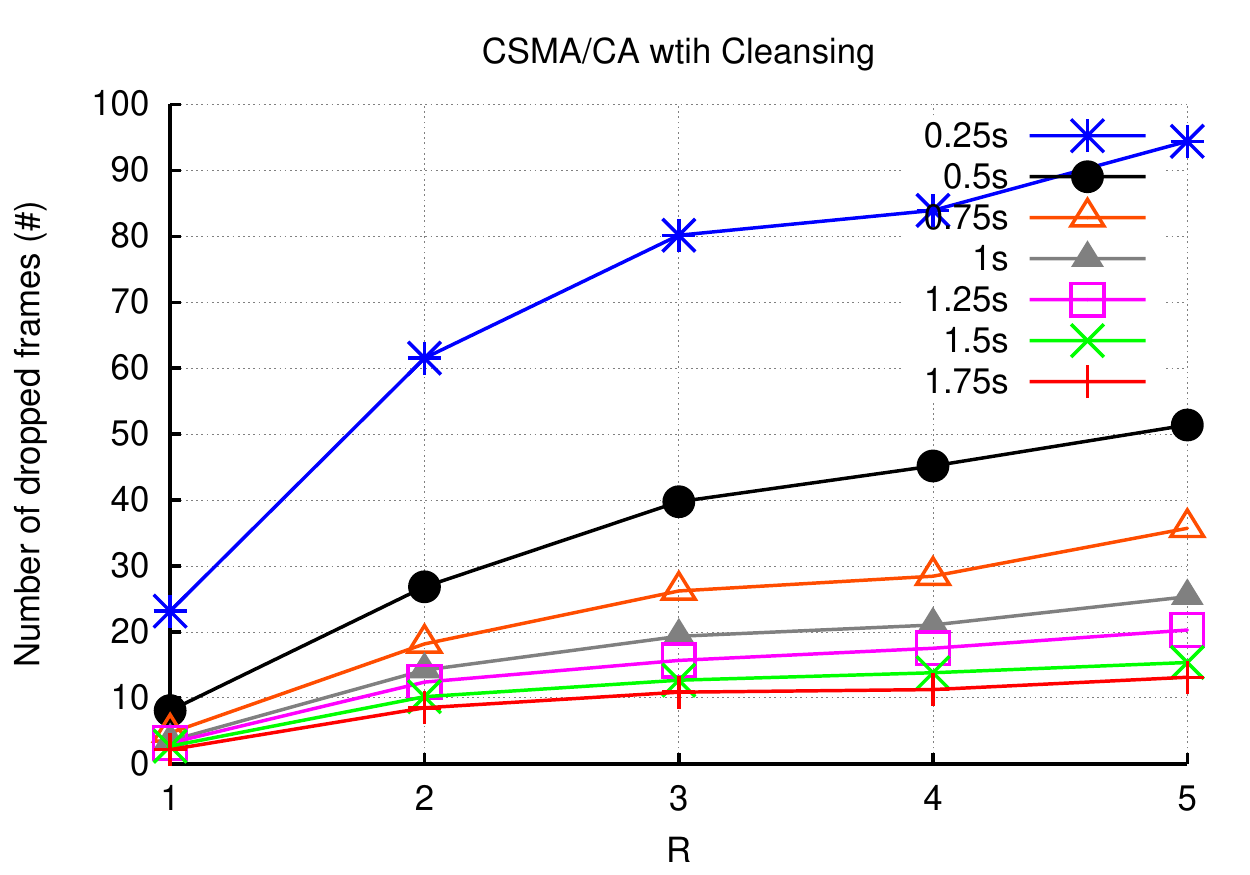}
  }
  \subfloat[Frame queue time]{
    \centering
  	\label{fig:gridqueue-cleans}
  	\includegraphics[width=0.3\textwidth]{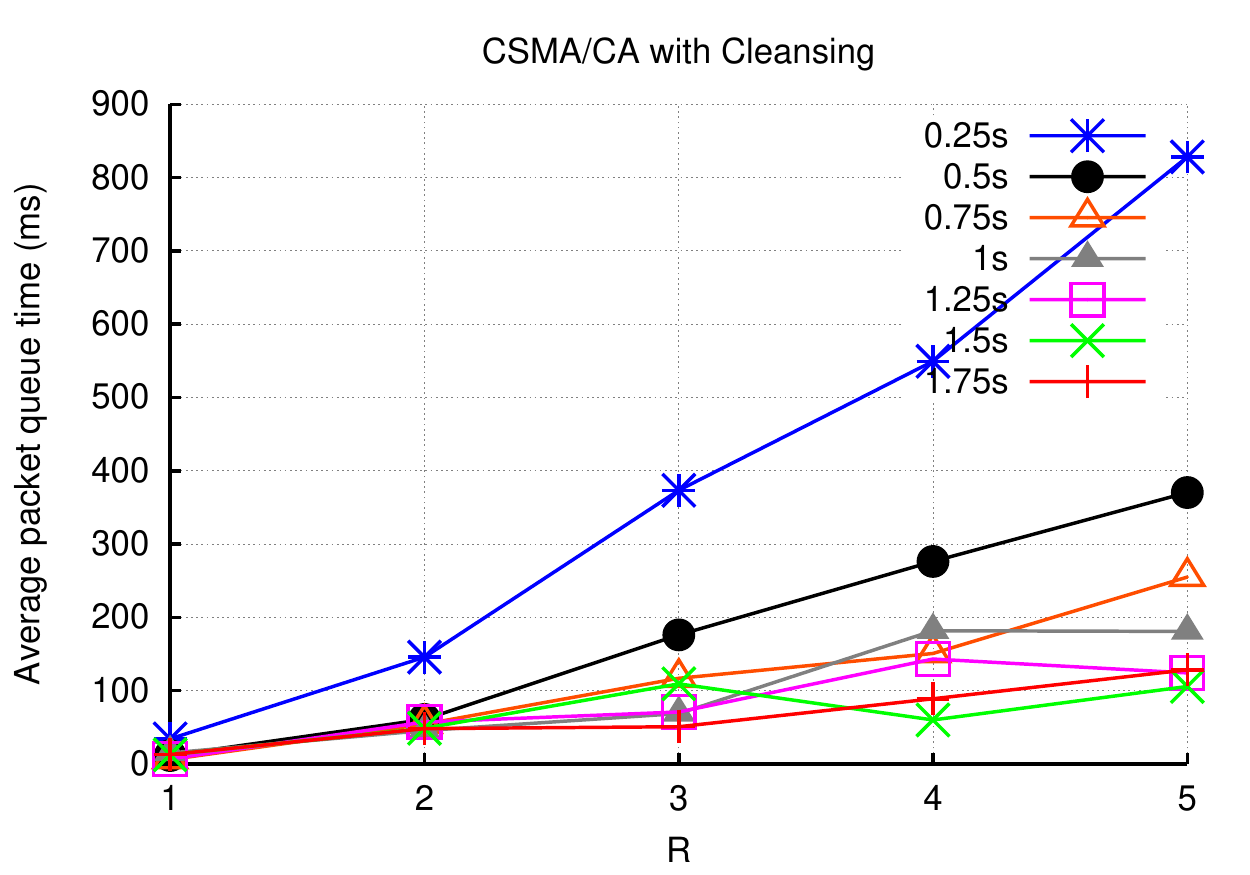}
  }
  \caption{Average number of transmissions, retransmissions and average frame queue time in the grid scenario, with (d-f) and without (a-c) MAC Cleansing, for different values of $I_{\min}$, $k=1$ and $\eta=1/2$.}
    \vspace{-1.5em}
  \label{fig:tx}
  \label{fig:tx-cleans}

\end{figure*}

Figure~\ref{fig:tx} shows the average number of transmissions and retransmissions during the entire simulation. As the range of each node grows, fewer messages are required to cover the entire network. Trickle then performs well, suppressing many transmissions (Figure~\ref{fig:gridtx}). However, many of the messages are actual retransmissions from the MAC layer (Figure~\ref{fig:gridrtx}). Since $k=1$, these are obsolete messages. Furthermore, due to the congested media, frames are left in the queue for a longer time (Figure~\ref{fig:gridqueue}), often leading to chained attempts for retransmission and further back-offs.

Figures \ref{fig:gridtx-cleans}-\ref{fig:gridqueue-cleans} show the impact of using Cleansing. CSMA/CA with Cleansing is aggressive with cleaning the MAC queue, as is visible in Figure~\ref{fig:griddrp-cleans}. This makes Trickle work as intended even for small values of $I_{\min}$. Additionally, the average queue time is considerably lower compared to the original CSMA/CA.

\subsection{Hardware experiments}
\begin{figure*}[ht]
\centering
\vspace{-2.0em}
\subfloat[Physical layout]{
\label{fig:iotlab-layout} \includegraphics[width=0.38\textwidth]{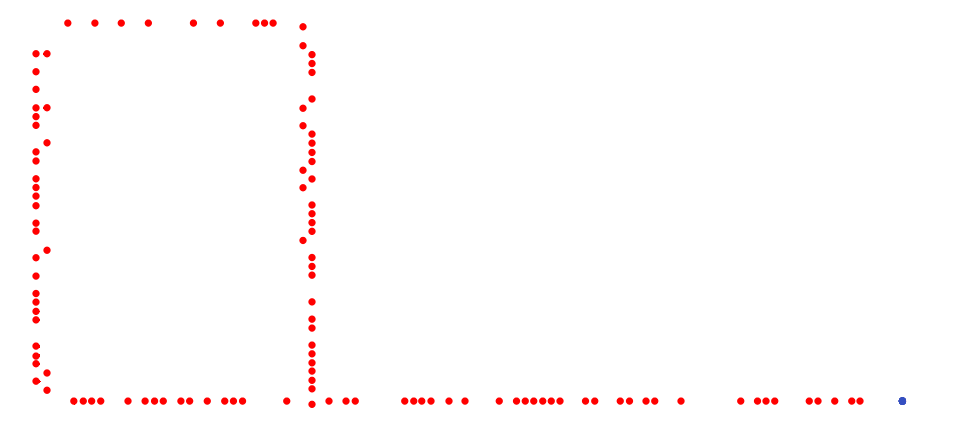}
}
\subfloat[Delay]{
\label{fig:iotlab-delay} \includegraphics[width=0.38\textwidth]{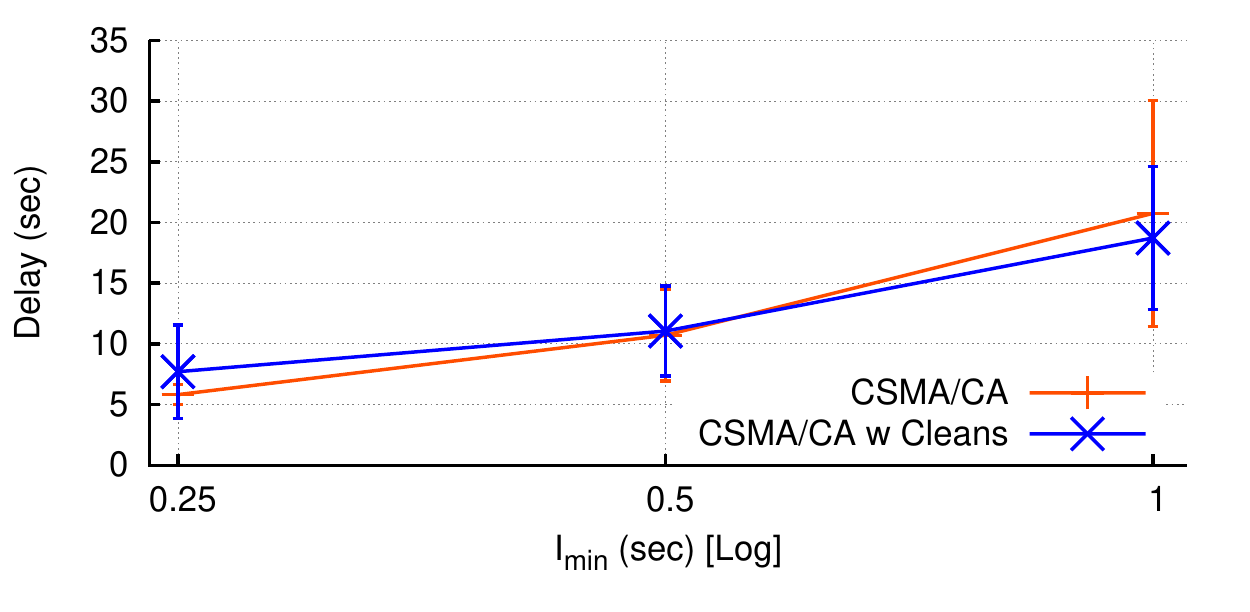}
} \\   \vspace{-1.0em}
\subfloat[Transmissions]{
		\label{fig:iotlab-tx} \includegraphics[width=0.38\textwidth]{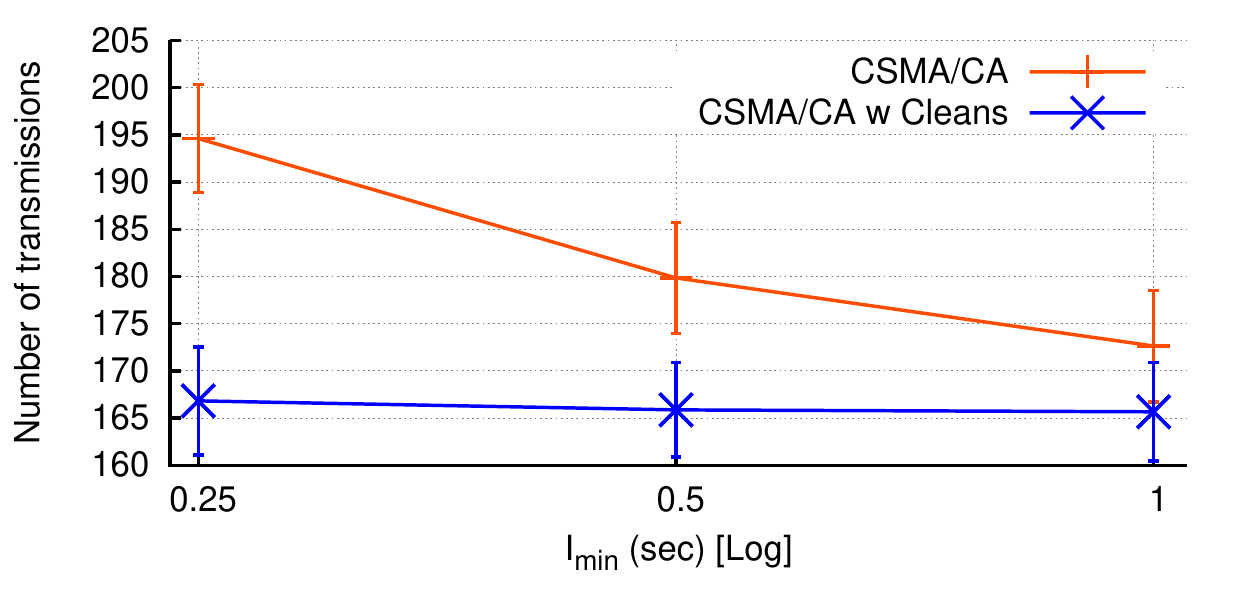}
}
\subfloat[Retransmissions]{
		\label{fig:iotlab-rtx} \includegraphics[width=0.38\textwidth]{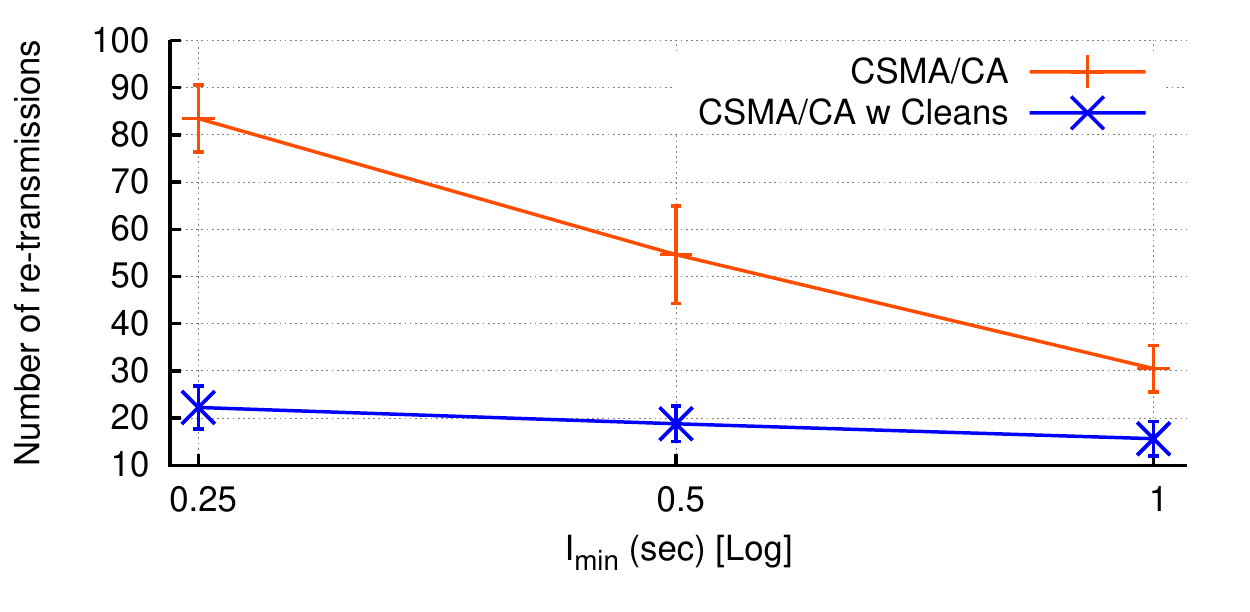}
}\vspace{-0.5em}
\caption{Experimental results from the IoT-Lab test bed. An update is injected at the bottom-right node, and is propagated using Trickle. The entire network is reachable in 12 hops. We show the  averages  and standard deviations over 100 executions.}
\vspace{-1em}
  \label{fig:iotlab}
\end{figure*}

To confirm the simulation results, we ran the same application on a physical test bed provided by FIT IoT-LAB~\footnote{http://www.iot-lab.info}. The test bed consists of 119 STM32 (ARM Cortex M3) based nodes, with the AT86RF231 IEEE 802.15.4 radio chip, arranged as in Figure~\ref{fig:iotlab-layout}. As before, all nodes use the ContikiMAC RDC protocol with a wake-up frequency of 8 Hz. The redundancy constant was fixed to $k=1$, with $I_{\min}$ set to 0.25s, 0.5s and 1.0s.  For each setting, we ran 100 executions of Trickle dissemination of one update, injected at the bottom-right node.

Figure~\ref{fig:iotlab-rtx} shows that using CSMA/CA, low values of $I_{\min}$ introduce a lot of collisions, which force retransmissions by the MAC layer. Increasing $I_{\min}$ helps reduce the number of transmissions (Figure~\ref{fig:iotlab-tx}), but at the expense of a higher delay (Figure~\ref{fig:iotlab-delay}). On the other hand, CSMA/CA with Cleansing has consistent performance using all three different values of $I_{\min}$. Due to the proactive purging policy, the number of messages remains comparable with different values of $I_{\min}$. As expected, the delay increases together with $I_{\min}$, but it is still in the same range as with the original CSMA/CA.

\section{Conclusion}\label{sec:conclusion}
In this paper we analyzed the performance of the Trickle algorithm for data dissemination when used in combination with low-power MAC protocols. We analyzed how the interplay of radio duty cycling and CSMA back-offs can contribute to bad Trickle performance. Analytically, we showed the MAC layer introduces inconsistencies, which lead to redundant transmissions and large update delays.

In order to resolve these issues, we proposed a small modification to the MAC layer, called \textit{Cleansing}. The Cleansing MAC modification purges obsolete Trickle messages that are sent due to the inconsistencies caused by the MAC layer.

Through a simulation study, and then confirmed with experiments on a large physical test bed, we showed that the Cleansing MAC indeed improves performance. We found that the number of redundant transmissions in dense topologies is decreased greatly and that the update speed in networks with bottlenecks is improved drastically.

As future work, we plan to extend the analysis to environments where the redundancy constant is greater than one. Additionally, we want to generalize the impact of the data link layer to Trickle timing, regardless of the combination of MAC protocol and radio duty cycling protocol.

\section*{Acknowledgments}
The authors would like to thank Onno Boxma for the many useful comments during the writing of this text. This work is supported in part by the Dutch P08 SenSafety Project, as part of the COMMIT program.

\bibliographystyle{splncs03}
\bibliography{bibliography,rfc}

\appendix
\section{Calculation of $\mathbb{P}^{\text{bo}}_{n,b}$ and $\mathbb{E}[N_n^r]$}\label{app}
Assume we have a network of $n$ synchronized nodes all within communication distance of each other, running Trickle-based dissemination, with $k=1$ and $\eta=1/2$. We are interested in calculating the probability that $b$ nodes will schedule a CSMA back-off during a single interval of length $I_{\min}=m \cdot w$.

Denote by $t_1$ the time that the first node will start broadcasting. Looking at a single other node, it will receive the packet at some time $t_r$ uniformly in $[t_1,t_1+w]$. Note that it is possible that $t_r>I_{\min}$, i.e. reception of a packet occurs outside of the interval in which it was scheduled. Denote by $t_2$ this node's broadcasting time. This node will schedule a back-off if it picked its broadcasting time in the interval $[t_1,t_r]$. Similarly, the node will suppress its transmission if it picked its broadcasting time after time $t_r$. Hence, taking into account all possible combinations of $b$ nodes scheduling a back-off and $n-b-1$ that do not, we arrive at:
\[ \mathbb{P}^{\text{bo}}_{n,b}:=n \binom{n-1}{b} \mathbb{P}\left[t_2\in[t_1,t_r]\right]^b\mathbb{P}\left[t_r\leq t_2\right]^{n-b-1}\]
 \begin{equation}\label{pnb}= n \binom{n-1}{b}\int\limits_{I_{\min}/2}^{I_{\min}}\mathbb{P}\left[t_2\in[t_1,t_r] \text{ }\vert\text{ }t_1=t\right]^b\mathbb{P}\left[t_r\leq t_2\text{ }\vert\text{ }t_1=t \right]^{n-b-1}\text{d}\mathbb{P}[t_1\leq t].
\end{equation}
Recall that both $t_1$ and $t_2$ are picked uniformly in $[I_{\min}/2,I_{\min}]$. Hence,
\begin{equation}
\mathbb{P}\left[t_2\in[t_1,t_r] \text{ }\vert\text{ }t_1=t\right]=\frac{2}{I_{\min}}\int\limits_{t}^{t+w}\frac{1}{w}\int\limits_{t_1}^{\min\left[t_r,I_{min}\right]}\text{d}t_2\text{ d}t_r,
\end{equation}
\begin{equation}
\mathbb{P}\left[t_r\leq t_2 \text{ }\vert\text{ }t_1=t\right]=\frac{2}{I_{\min}}\int\limits_{t}^{t+w}\frac{1}{w}\int\limits_{\min\left[t_r,I_{min}\right]}^{I_{\min}}\text{d}t_2\text{ d}t_r.
\end{equation}
Distinguishing between $t_1 \leq I_{\min}-w$ and $t_1> I_{\min}-w$ and paying careful attention to the minima in the integral limits, after substitution Equation \eqref{pnb} becomes:
\begin{equation}\label{pnbsplit}
n\binom{n-1}{b}\left(\frac{2}{I_{\min}}\right)^n\int\limits_{I_{\min}/2}^{I_{\min}-w} \left(\frac{w}{2}\right)^b\left(I_{\min}-t_1-\frac{w}{2}\right)^{n-b-1}\text{ d}t_1 +
\end{equation}
\[
n\binom{n-1}{b}\left(\frac{2}{I_{\min}}\right)^n\int\limits_{I_{\min}-w}^{I_{\min}} \left(\frac{I_{\min}-t_1}{2w}\left(2w-I_{
\min}+t_1\right)\right)^b\left(\frac{(I_{\min}-t_1)^2}{2w}\right)^{n-b-1}\text{d}t_1.
\]
Calculating the first integral of \eqref{pnbsplit}, we find:
\begin{equation}\label{pnbsplit1}
\binom{n}{b}\frac{(m-1)^{n-b}-1}{m^n}.
\end{equation}
Substituting $z=(I_{\min}-t_1)/(2w)$ simplifies the second integral to the well-known incomplete Beta function:
\begin{equation}\label{pnbsplit2}
n\binom{n-1}{b}\left(\frac{4}{m}\right)^n\int_{0}^{\frac{1}{2}}(1-z)^b z^{2n-b-2}\text{ d}z.
\end{equation}
Combining Equations \eqref{pnbsplit1} and \eqref{pnbsplit2}, we find:
\begin{equation}\label{pnbfinal}
\mathbb{P}_{n,b}=\binom{n}{b}\frac{(m-1)^{n-b}-1}{m^n}+n\binom{n-1}{b}\left(\frac{4}{m}\right)^n\int_{0}^{\frac{1}{2}}(1-z)^b z^{2n-b-2}\text{ d}z.
\end{equation}
For $b=0$ this simplifies to:
\begin{equation}
\mathbb{P}^{\text{bo}}_{n,0}= \frac{1}{m^n}\left((m-1)^n+\frac{1}{2n-1}\right).
\end{equation}
Finally, we can use \eqref{pnbfinal} to calculate the expected number of back-offs. Switching the order of summation and integration, we derive:
\begin{equation}\label{expec}
\mathbb{E}[N^r_n]:=\sum_{i=0}^n i\mathbb{P}_{n,i}=\frac{n}{m}-\frac{1}{n+1}\left(\frac{2}{m}\right)^n.
\end{equation}


\end{document}